\pdfoutput=1
\documentclass{article}
\usepackage{graphicx, indentfirst, hyperref, natbib, enumitem}
\usepackage[a4paper, margin = 3cm]{geometry}
\makeatletter
\let\mailmark\@fnsymbol
\makeatother

\newcommand*{\cf}{\emph{cf.}}
\newcommand*{\eg}{\emph{eg.}}
\newcommand*{\etc}{\emph{etc}}
\newcommand*{\prog}[1]{\emph{#1}}
\newcommand*{\tblref}[1]{Table \ref{tbl:#1}}
\newcommand*{\figref}[1]{Figure \ref{fig:#1}}
\newcommand*{\secref}[1]{Section \ref{sec:#1}}
\newcommand*{\tblfn}[1]{\thxmark{\mailmark{#1}}}
\let\thxmark\textsuperscript
\let\cite\citep

\begin{document}
\title{%
	Detector integration at HEPS:\\
	a systematic, efficient and high-performance approach%
}
\author{%
	Qun Zhang\thxmark{1,2}, Peng-Cheng Li\thxmark{1,3,4},
	Ling-Zhu Bian\thxmark{1}, Chun Li\thxmark{1},\\
	Zong-Yang Yue\thxmark{1}, Cheng-Long Zhang\thxmark{1},
	Zhuo-Feng Zhao\thxmark{2},\\Yi Zhang\thxmark{1,4}, Gang Li\thxmark{1},
	Ai-Yu Zhou\thxmark{1}, Yu Liu\thxmark{1,\mailmark{1}}%
}
\date{}
\maketitle
\begingroup
\renewcommand{\thefootnote}{\fnsymbol{footnote}}
\footnotetext[1]{\ Correspondence e-mail: \texttt{liuyu91@ihep.ac.cn}.}
\endgroup
\footnotetext[1]{\ %
	Institute of High Energy Physics, Chinese Academy of Sciences,
	Beijing 100049, People's Republic of China.%
}
\footnotetext[2]{\ %
	North China University of Technology,
	Beijing 100144, People's Republic of China.%
}
\footnotetext[3]{\ %
	National Synchrotron Radiation Laboratory,
	University of Science and Technology of China,
	Hefei, Anhui 230029, People's Republic of China.%
}
\footnotetext[4]{\ %
	University of Chinese Academy of Sciences,
	Beijing 100049, People's Republic of China.%
}

\section*{Synopsis}

Keywords: detector integration, GenICam,
high-performance readout, Python IOC, software architecture.

To integrate the at least 25 kinds of detector-like devices to be used
at HEPS Phase I, a systematic and efficient strategy is applied.  A
careful separation of concerns between collaborating groups of personnel
is followed to boost efficiency, and an extended \prog{ADGenICam} is
used to reduce code repetition in \prog{EPICS} IOCS for detectors.
An \prog{areaDetector} workalike in the \prog{QueueIOC} Python IOC
framework is used to integrate high-performance detectors.

\section*{Abstract}

At least 25 kinds of detector-like devices need to be integrated in Phase I
of the High Energy Photon Source (HEPS), and the work needs to be carefully
planned to maximise productivity with highly limited human resources.  After
a systematic analysis on the actual work involved in detector integration,
a separation of concerns between collaborating groups of personnel is
established to minimise the duplication of efforts.  To facilitate software
development for detector integration, the \prog{ADGenICam} library, which
abstracts repeated code in \prog{EPICS} modules for cameras, is extended to
support a much wider range of detectors.  An increasingly considerable fraction
of detectors, both inside and outside HEPS, offer performance that exceed
capabilities of the \prog{areaDetector} framework in \prog{EPICS}.  Given this
background, \prog{areaDetector}'s limitations in performance and architecture
are analysed, and a \prog{QueueIOC}-based framework that overcomes these
limitations is introduced.  A simple, flexible \prog{ZeroMQ}-based protocol
is used for data transport in this framework, while RDMA transport and
multi-node readout will be explored for higher data throughputs.  By calling
C/C++ libraries from within Python, the performance of the former and the
expressiveness of the latter can coexist nicely; the expressiveness allows
for much higher efficiency in the implementation and use of integration
modules functionally comparable to their \prog{EPICS} counterparts.

\section{Introduction}\label{sec:intro}

In recent years, progress is being rapidly made in the research and
construction of advanced light sources, \eg\ 4th-generation synchrotrons
and free-electron lasers.  While pushing towards the physical limits, the
excellent attributes of these light sources -- small light spots, high
brightness, strong coherence \etc\ -- allow scientists and engineers to
do researches in much finer spatial and temporal scales.  Correspondingly,
radiation detection techniques \cite{fajardo2017} are also advancing rapidly,
providing instrumental bases for practical exploitation of the attributes
above.  Improvements on the pixel sizes (in combination with the numbers of
pixels) and framerates of detectors enable exploration of matter on finer
spatial and temporal scales, respectively.  Higher framerates also allow
for the use of techniques like fly scans \cite{li2023a, li2023b}, which
greatly reduce the time spent on scanning which increases because of the
smaller light spots and the more scan points; these techniques also help
to reduce the radiation damage in samples because of the higher brightness.

Meanwhile, we have noticed that the mainstream software frameworks used
by major light sources around the world may have themselves become a main
bottleneck in the application of high-performance detectors.  Both the
Beijing Synchrotron Radiation Facility (BSRF) and the High Energy Photon
Source \cite[HEPS, \cf][]{xu2023} use the Experimental Physics and
Industrial Control System (\prog{EPICS}) as the main environment for device
control; the main framework for \prog{EPICS}-based detector integration is
\prog{areaDetector} \cite{rivers2017}, which is the main example in our
discussion.  According to our experience, when writing HDF5 files using
\prog{areaDetector}'s \prog{HDF5Plugin}, the \emph{data throughput} has
an upper bound of around 500-600\,MB/s.  In contrast, even only among
the requirements of detector integration from the 15 beamlines in
Phase I of HEPS (\tblref{perf-req}), there are already quite a few
detectors exceeding 1\,GB/s (some even exceeding 10\,GB/s) throughput,
each with some beamline(s) wanting to exploit its performance as much
as reasonable.  Outside BSRF and HEPS, readout requirements reaching
1\,GB/s or even 10\,GB/s are also becoming increasingly common
\cite{fischer2017, yendell2017, gories2023, dong2022}.

\begin{table}[htbp]\centering
\caption{%
	A few representative high-performance
	detectors that need to be integrated at HEPS.%
}\label{tbl:perf-req}\vspace*{1em}
\begin{tabular}{cccc}\hline
Detector model & Max.\ resolution & Max.\ framerate & Max.\ throughput \\\hline
Tucsen camera A & 6\,K$\times$6\,K$\times$2\,B & 19\,Hz & 1.3\,GB/s \\
Tucsen camera B\,\tblfn{1} & 5\,K$\times$4\,K$\times$1\,B & 450\,Hz & 8.8\,GB/s \\
A self-made Si-pixel detector &
	2$\times$5.76\,M$\times$2\,B\,\tblfn{2} & 1000\,Hz & 22\,GB/s \\
Eiger X 1M & 1\,K$\times$1\,K & 3000\,Hz & Maybe $>$1\,GB/s\,\tblfn{3} \\
Hamamatsu C15550-20UP &
	4096$\times$2304$\times$2\,B & 120\,Hz & 2.1\,GB/s \\\hline
\multicolumn{4}{l}{\tblfn{1} Still in production.
	\,\tblfn{2} With dual thresholds. \,\tblfn{3} (Even) after compression.}\\
\end{tabular}
\end{table}

Apart from the data throughput issue above, we have also found
\prog{areaDetector} \prog{HDF5Plugin} has an about 4\,kHz upper bound in
\emph{framerates} when writing HDF5 files; meanwhile, fly-scan experiments
on the 10\,kHz level \cite{deng2019, batey2022} are a field pursued actively
by us.  Even only among the requirements from HEPS Phase I, experiments
exceeding 10\,kHz at least include coherent diffraction imaging (CDI) and X-ray
photon correlation spectroscopy (XPCS) experiments at the hard X-ray coherent
scattering beamline (B4), as well as 2D real-space current mapping experiments
at the high-resolution nanoscale electronic structure spectroscopy beamline
(BC).  Apart from possibly demanding high-throughput readout, these
experiments may also demand data readout from sensors, motion controllers,
fly-scan controllers \cite[like PandABox, \cf][]{christian2019} \etc.
Although the latter kind of readout has high framerates, its throughputs
may not be very high; because of the smaller size of each data frame, the
devices that produce these data may be regarded as \emph{0D/1D detectors}
\cite{li2023b}.  Unfortunately, although \prog{areaDetector} supports
1D detectors like Mythen, it does not seem to consider 0D readout at
all, making it non-trivial to integrate 0D devices in a unified way.

Aside from the performance considerations above, another issue we notice is
the costs of integration, where the first problem is the number of detector
brands: even only at HEPS Phase I, we have at least 25 kinds of detector-like
devices to be integrated (\tblref{type-req}), and we believe similar problems
will also occur at other light sources, especially new light sources.  Only
about a half among these devices have open-source \prog{EPICS} modules
(input/output controllers or IOCs), and in our experience a majority of
these modules also need some customisations or even fixes.  Among the other
half, most require us to write high-performance integration modules, while
the rest need self-developed \prog{EPICS} IOCs.  During the still ongoing
process of integrating these devices, we have realised that a large fraction
of the costs in development, maintenance, application and learning of
their \prog{EPICS} IOCs come from \prog{EPICS} itself, whose programming
interfaces result in significant amounts of \emph{redundant code} that
are hard to eliminate.  From our quite crude research on \prog{TANGO}'s
detector integration framework, \prog{LImA} \cite{homs2011}, the
latter has similar code redundancy problems to some extent.

\begin{table}[htbp]\centering
\caption{Main detector-like devices that need to be integrated at HEPS Phase I.}
\label{tbl:type-req}\vspace*{1em}
\begin{minipage}[t]{0.9\textwidth}
\begin{itemize}[itemsep = 0pt]
\item Devices with \prog{EPICS} IOCs that satisfy current requirements:
\begin{itemize}[nosep]
	\item GenICam (GigE/USB3 Vision) industrial cameras: Hikvision, FLIR \etc
	\item Andor (sCMOS, CCD), marCCD, Merlin, Minipix, PICam, Pilatus
	\item Keck-PAD, PCO, Photron, PVCAM; Mythen; Falcon Xn, Xspress 3
\end{itemize}
\item Devices that require self-developed \prog{EPICS} IOCs:
\begin{itemize}[nosep]
	\item Specialized Imaging Kirana, Ximea
	\item Self-made APD detector, iRay Mercu
\end{itemize}
\item Devices that require high-performance integration modules:
\begin{itemize}[nosep]
	\item Eiger, Lambda/Sparta, Rigaku XSPA, Hamamatsu
	\item Tucsen, self-made Si-pixel detector
	\item PandABox, multiple sensors (needing high-speed 0D readout)
\end{itemize}
\end{itemize}
\end{minipage}
\end{table}

With the problems above in mind, we created the \prog{QDetectorIOC}
framework for detector integration based on our Python IOC framework
\prog{QueueIOC} \cite{li2024}, with performance, versatility and
succinctness as its major goals; all Python IOC examples in this paper
are available from the \prog{QueueIOC} code repository.  However, before
the exposition of \prog{QDetectorIOC} in \secref{qdetector}, in this paper
we will also cover the separation of concerns in detector integration
established at HEPS in \secref{personnel}, and our modifications to
the \prog{ADGenICam} module in \prog{areaDetector} that vastly extend its
range of supported devices in \secref{genicam}.  In addition to being
informative in themselves, these preceding sections also convey notions
and methods which have underlain the foundation of \prog{QDetectorIOC}.

\section{Separation of concerns in detector integration at HEPS}
\label{sec:personnel}

The number of detector brands mentioned in \secref{intro} is far from the
only source of complexity in detector integration at HEPS; another important
source of complexity is the diversity of these detectors.  In terms of
hardware mechanisms, the detectors at HEPS cover direct/indirect detection,
Si/Ge/GaAs sensors, CCD/sCMOS, position/energy/time resolution, \etc;
in terms of manufacturers, they include self-made detectors and commercial
detectors; in terms of application scenarios, they are used in common types
of experiments like imaging, spectroscopy, diffraction and scattering.
The readout requirements cover 0D to 2D data frames; devices may also
attach metadata, \eg\ PandABox's ``table headers'', which are essential for
downstream data processing.  To facilitate parallelised processing from the
origin, high-throughput detectors may also involve requirements like frame
splitting and multi-stream readout; tomography experiments may pose further
requirements on the orientation of frame splitting, in order to accommodate
parallelisation of the inverse Radon transform \cite{weisstein2024}.
The complexity in the details above lead to the strong \emph{diversity
in the software/hardware interfaces of detectors}, which is a most
prominent difficulty in detector integration at HEPS and perhaps most
other light sources.  Consequently, a most important role of frameworks like
\prog{areaDetector}, from our perspective, is an adapter that converts these
diverse interfaces into some kind of unified interfaces, as much as reasonable.

While it is obvious from the above that integration of detector-like devices is
a complicated systematic task, first-hand experience shows that an overwhelming
majority of time and energy in this task is spent on writing and debugging
\emph{integration modules}.  Therefore when exploring for an efficient
way to (unified) detector integration, it is instructive to begin with the
observation of the code of these modules.  By differentiating between parts
shared between modules and parts unique to each module, we may identify those
procedures in detector integration that are closely tied to the detector in
question, and simplify repeated work in the rest procedures.  Insightful
observations of the modules can also reveal jobs that can be better handled by
other groups of personnel, and help us to form a reasonable \emph{separation
of concerns} between collaborating groups; in this way we can focus on
writing and debugging integration modules, thus maximising the collective
productivity of these groups.  During the work on integration modules, which
for us at BSRF/HEPS are mainly \prog{EPICS} IOCs, we came to the conclusion
that the device-specific code in these modules roughly fall into two
categories: \emph{control communication} and \emph{data communication}
with devices; the former does device (un)initialisation and configuration
of parameters, while that latter is responsible for data readout.

While readout interfaces (including the formats of raw data) can vary
greatly between different detectors, what the raw data describe are almost
invariably \emph{multidimensional arrays} composed of a number of frames with
stable lengths (usually fixed at least throughout a readout session): \eg\ a
3D array composed of 100 frames, each being a 1080$\times$1920 \verb|uint16|
array.  At HEPS and many other light sources, the data are first saved
into HDF5 files, then they are assembled into the NeXus format and get
postprocessed.  The requirements of data postprocessing (in the general
sense), including the saving, assembling, visualisation, reduction \etc\ %
of data, are highly diverse as they depend on factors like the detector
type, the experiment type \etc.  Nevertheless, these tasks are not
directly reflected in the code of integration modules, and therefore may
be considered job of the group dedicated to postprocessing; at HEPS, the
\prog{Mamba} Data Worker (\prog{MDW}) framework \cite{li2023c} is responsible
for the latter job.  Given these considerations, we define the boundary
between us and the postprocessing group as some unified data protocols that
transmit multidimensional arrays, whether HDF5 files or the protocol in
\secref{qdetector} based on \prog{ZeroMQ} or RDMA.  We are also responsible
for providing essential facilities demanded by data postprocessing, \eg\ %
the support for frame splitting and multi-stream readout in \secref{intro}.

\begin{figure}[htbp]\centering
\includegraphics[width = 0.8\textwidth]{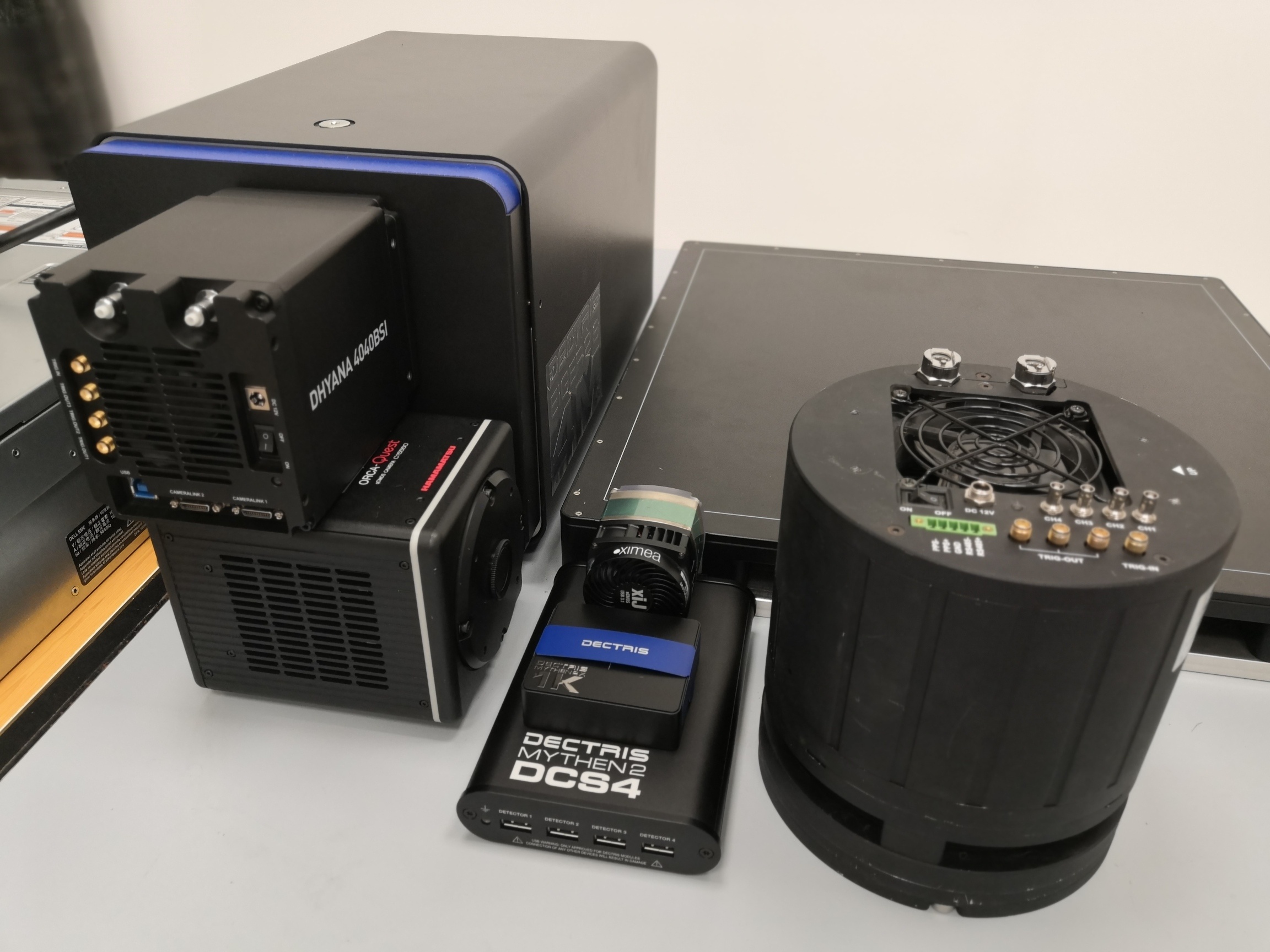}
\caption{%
	One batch test for detectors and their integration modules at HEPS
	involved an Eiger2 S 4M, a Mythen2 X 1K, a Ximea xiJ, a Hamamatsu
	C15550-20UP, a Tucsen 6\,K$\times$6\,K, a Tucsen 4\,K$\times$4\,K
	and an iRay Mercu flat panel.  These particular devices were provided
	by the engineering materials beamline (B1), the hard X-ray nanoprobe
	multimodal imaging beamline (B2), the hard X-ray high-resolution
	spectroscopy beamline (B5), the hard X-ray imaging beamline (B7)
	and the pink-beam SAXS beamline (BB); but of course, other beamlines
	that use similar detectors will also benefit from this test.%
}\label{fig:batch-test}
\end{figure}

As most detector-like devices do not provide separable vendor interfaces for
control communication and data communication, in addition to data readout
integration modules usually also need to do device control.  The diversity
in the types and uses of detectors are mainly reflected in the control
communication (and postprocessing, as are discussed above).  Greatly adding
to the complexity in control communication is the numerous \emph{control
parameters} of most detectors: \eg\ each industrial camera usually has
more than 100 parameters.  To avoid an unlimited growth of workload, when
writing and debugging integration modules it is obviously necessary to
establish some kind of precedence for the different aspects of a device's
interface.  We have chosen to first care about the common, essential
aspects, like triggering and acquisition modes (including necessary
hardware aspects, \eg\ electronic/electrical issues like TTL/LVDS wiring
and safe voltages), with achieving simple and reliable data readout as
the most important goal.  After an integration module passes our basic
tests (\eg\ the batch test in \figref{batch-test}), we deliver it to
beamlines, and help beamline scientists confirm whether the control
parameters and hardware properties meet their demands; of course,
with the growth of our experience, we may proactively test more
aspects of a device's interface when our schedule permits.

\section{\prog{EPICS}-based integration and an extended \prog{ADGenICam}}
\label{sec:genicam}

With the separation of concerns in \secref{personnel}, we are able to
concentrate on writing and debugging integration modules for detector-like
devices at HEPS.  However, even with this it would be a great burden to
integrate these devices (\cf\ \tblref{type-req}), if the traditional way
to deploy \prog{EPICS} modules was employed without automation.  Fortunately,
the \prog{ihep-pkg} packaging system and the \verb|~/iocBoot| convention
\cite{liu2022, li2024} have been created to simplify \prog{EPICS} deployment
by eliminating or abstracting repeated steps in building and configuring
\prog{EPICS} modules; a software/hardware handbook is also provided to
provide simple, reproducible instructions for basic configurations of the
controlled devices.  With \prog{ihep-pkg}, we are also able to encode our
customisations or fixes to device IOCs (\cf\ \secref{intro}) as patches
that get automatically applied in the packaging process, so that a packaged
IOC from \prog{ihep-pkg} is in a reproducibly usable state out of the box.
Vendor-provided API libraries, conventionally called ``\emph{SDKs}'' (software
development kits), are also packaged in \prog{ihep-pkg}.  The packaging of
SDKs not only reduces the workload in installation, but also helps us to
improve reproducibility in SDK behaviours because of the better control
of SDK versions and certain subtle installation steps; it also enables
convenience like the unified use of user group \verb|video| under
Linux to grant access to devices using USB or kernel drivers.

Following the same line of thought, it is also natural for us to attempt
to integrate multiple kinds of devices with a unified set of IOCs, saving a
considerable amount of programming workload: \eg\ use of the \prog{ADAravis}
IOC, which is based on the open-source \prog{Aravis} SDK, to integrate
most GigE/USB3 vision industrial cameras.  The rest of this section covers
how we modify \prog{ADGenICam}, so that it can support devices like those
from Ximea, which are unsupported by the former's original version; similar
mechanisms are also used to integrate devices from Hamamatsu, Tucsen
\etc\ in \secref{qdetector}.  With these measures taken, we are hopefully
able to integrate all devices in \tblref{type-req} by the end of 2025
with less than a handful of people which can only work part-time on device
integration.  The point in using \prog{ADGenICam} is to simplify IOCs based
on it by systematically reducing code repetition, from exploiting what we call
the \emph{``BCDEIS'' type system} which GenICam-based cameras comply with.
The SDK of a compliant device exposes a few functions, each used to read/write
control parameters (``\emph{features}'') in a given type: boolean, command,
float/double, enumerator, integer or string, where the feature ID is passed
as an argument to these functions; our abbreviation for each type was taken
from the files \verb|makeDb.py| and \verb|ADGenICam.cpp| in \prog{ADGenICam}.

The code generator \verb|makeDb.py| creates \prog{EPICS} databases (templates)
from XML descriptions of cameras, encapsulating the feature names in the
\verb|@asyn| string; in this way, \prog{ADGenICam}-based IOCs are able to
deal with devices with relatively dynamic feature sets.  The latter string
is parsed by the \verb|drvUserCreate()| function in \prog{ADGenICam} when
\prog{EPICS} process variables (PVs) are created according to the databases,
mapping each PV to the corresponding feature name and type.  The actual mapping
from the abstract BCDEIS type system to concrete SDK functions is done in
IOC source code files for subclasses of the \verb|GenICamFeature| class
from \prog{ADGenICam}.  The problem with attempts to integrate devices from
Ximea, Hamamatsu, Tucsen \etc\ based on \prog{ADGenICam} is that even though
their SDKs may be based on GenICam-based lower-level interfaces, they do not
directly expose these interfaces.  For example, with the \prog{harvesters}
(\url{https://github.com/genicam/harvesters}) library officially recommended
for GenICam, we are able to extract XML descriptions from Ximea devices
\cite[\cf\ \texttt{makeXml.py} from our modified version of \prog{ADGenICam},
available from the open-source edition of \prog{ihep-pkg},][]{li2024}.
However, these XML files use feature names like \verb|AcquisitionMode|, which
differ from the feature names exposed by the Ximea SDK.  Nevertheless, from
the sketch above, it is not hard to realise that the mechanism actually does
not demand the feature names in SDK function calls to be strictly identical to
the feature names in an XML description.  Actually, it does not even require
SDK functions to use strings as feature IDs: \eg\ Hamamatsu devices use
integers, and a more complex example can be seen in \secref{qdetector}.

\begin{figure}[htbp]\centering
\includegraphics[width = 0.8\textwidth]{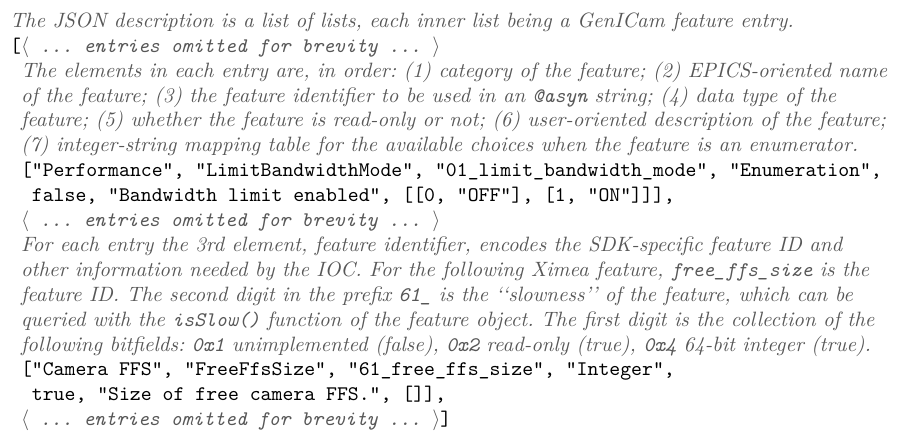}
\caption{%
	An excerpt from the JSON description file we created for a Ximea device.%
}\label{fig:gc-json}
\end{figure}

The feature sets for Ximea, Hamamatsu, Tucsen \etc\ are given in forms others
than XML descriptions: C/C++ headers, Python libraries, HTML/PDF documentation,
SDK functions that iterate through the features, \etc.  Consequently, instead
of using XML exporters, we need to write SDK-specific ``\emph{crawlers}''
(\eg\ \verb|ximeaJson.py| from the \prog{ADXimea} IOC to be covered below)
for feature sets, or in rare cases even transcribe the sets manually.  To
facilitate the process, we designed a JSON-based format (\figref{gc-json};
\cf\ also \figref{tucam-cffi}) for feature sets, which encodes all information
necessary for \verb|makeDb.py| and its companion \verb|makeAdl.py|; both
latter scripts have been modified to accept our JSON-based format in addition
to the original XML-based format.  As some code in the original version of
the class \verb|ADGenICam| from the \prog{ADGenICam} module conflicts with
the generalisation above, we have modified the module to provide a class
\verb|ADGenICamBase| which our IOCs, \eg\ \prog{ADXimea} for Ximea devices
(available from the open-source edition of \prog{ihep-pkg}), are based on;
\verb|ADGenICam| has been refactored based on \verb|ADGenICamBase| for full
backward compatibility.  To facilitate the polling of information like
sensor temperatures, as well as the refreshing of the entire feature set
when the value of one feature has been modified, also provided in our version
of \prog{ADGenICam} is a class \verb|ADGenICamSlow|.  This class automatically
updates certain subsets of the feature set upon given conditions, according
to the return value of the member function \verb|isSlow()| of each feature
object (\cf\ \figref{gc-json}): update periodically if the value is 0,
update upon feature modification if the value is 1, \etc.  Now with these
modifications, \prog{ADGenICam} can comfortably support most \emph{GenICam-like
detectors}, which is a great extension to the original range of quite strictly
GenICam-compliant cameras.  In principle quite a number of existent IOCs,
\eg\ \prog{ADAndor3} and \prog{ADPICam}, could also be rewritten into
\prog{ADGenICam}-based IOCs; but we have not yet done this actually,
as they are already quite usable, and we have a very tight schedule.

\section{Python-based integration using \prog{QDetectorIOC}}
\label{sec:qdetector}

Following the line of thought in \secref{personnel}, when discussing
\prog{QDetectorIOC}, our Python IOC framework for detectors, we first
focus on the control logic, and then discuss the data protocol.  As
the Tsuji (N)CT and the Ortec 974 counters are 0D detectors not used in
high-speed data readout at HEPS, their IOCs, respectively \verb|qdet_nct|
and \verb|qdet_o974|, use plain \prog{EPICS} PVs to output data.  Currently all
other QDetectorIOCs use the \prog{ZeroMQ}-based protocol to be covered later in
this section, so these two IOCs are good introductory examples for QDetectorIOCs
because of the simplicity of their data interface.  The next IOC in our brief
tour of the control logic in \prog{QDetectorIOC} is \verb|qdet_peak| for the
ARPES (angular-resolved photoemission spectroscopy) energy analyser from Scienta
Omicron, which demonstrates the use of a \emph{Python SDK}.  The device does
not yet produce a high throughput or framerate, but the lack of a C/C++ SDK
makes it ideal to integrate with a Python IOC.  \verb|helpers/qdet_peak_lib.py|,
a \emph{self-made SDK} (the self-made SDKs for other IOCs are at similar paths),
is used for this IOC instead of the PEAK SDK recommended by the vendor, for
the sake of succinctness and consistency with the rest of \prog{QDetectorIOC}.
Slightly more complex than \verb|qdet_peak| is \verb|qdet_eiger| for
Eiger, showcasing the application of \verb|QScanIOCBase|, the base class of
\verb|QScanIOC| \cite{li2024}, in QDetectorIOCs to implement periodic polling.
Even more complex is \verb|qdet_tucam| for Tucsen cameras, which feature
a mix of an old interface (\cf\ \url{https://github.com/djvine/ADTucsen})
with a GenICam-based interface.  Apart from the complexity of this
mixed interface, which is again encapsulated in a self-made SDK and
will be covered in the following paragraph, this IOC also shows how
a workalike of \verb|ADGenICamSlow| is used in \prog{QDetectorIOC}.

\begin{figure}[htbp]\centering
\includegraphics[width = 0.8\textwidth]{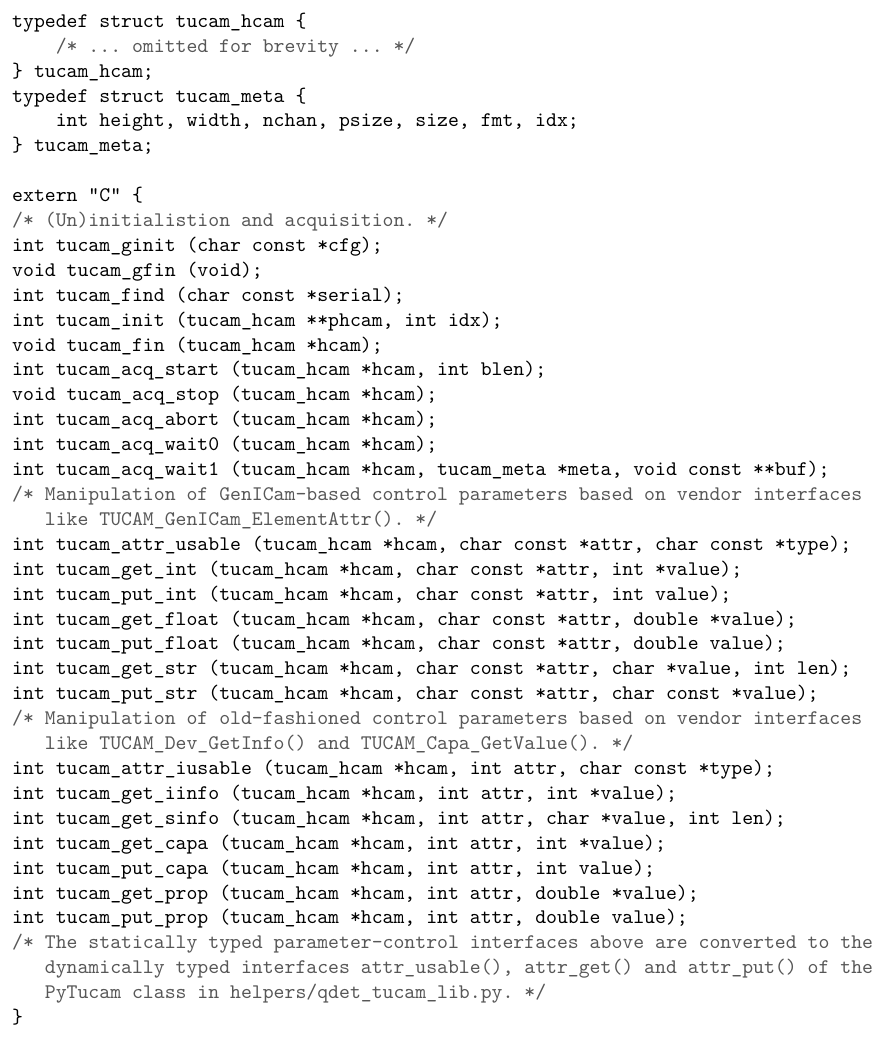}
\caption{%
	Interfaces provided by \texttt{helpers/qdet\string_tucam\string_lib.cc},
	the self-made C SDK for the \texttt{qdet\string_tucam}
	IOC, which are converted into Python interfaces in
	\texttt{helpers/qdet\string_tucam\string_lib.py};
	JSON crawlers for the IOC, \texttt{qdet\string_tucam\string_json.cc}
	and \texttt{qdet\string_tucam\string_json.py},
	are also available in the same subdirectory.%
}\label{fig:tucam-cffi}
\end{figure}

As the official SDKs for detector-like devices are mainly provided for C/C++,
in order for integration through Python IOCs, a majority of these devices
need self-made Python SDKs; the latter translate C/C++ vendor interfaces
into Python interfaces, by use of tools like \prog{ctypes}, \prog{cffi},
\prog{Cython} \etc.  A major and representative subset of them is GenICam-like
detectors, among which we find the Tucsen interface a most complex example;
however, the complexity of this interface comes not only from its mix of
old and new interfaces, but also from its quite bloated calling convention.
With careful encapsulation, we are able to create a succinct \emph{C ``SDK''}
(\figref{tucam-cffi}) that hides the latter kind of complexity; the latter SDK
also represents the ideal form of a detector SDK to us, both internally and
externally.  Internally, the SDK's implementation encapsulates the underlying
interface (a vendor SDK in the case of Tucsen) thinly to minimise performance
loss.  Externally, the SDK's interface is easily reusable and adaptable: \eg\ %
a performance test program for Tucsen cameras, \verb|tucamRate.c|, is given
in the supplementary materials, which is used to verify the cameras' maximal
framerates claimed by the vendor, and can be adapted for other GenICam-like
detectors after only minor changes.  The generality of the interface can also
be seen from a comparison between it and \verb|helpers/qdet_dcam_lib.c|,
our simplified C ``SDK'' for Hamamatsu, as the feature-related function
signatures from the latter are strongly reminiscent of those signatures
from the former that encapsulate Tucsen's old interface.  With similar
encapsulations, we are able to keep strong \emph{consistency} between our
self-made Python SDKs, and consequently strong consistency between the code
in corresponding Python IOCs, thus facilitating development and maintenance
of the code.  This also applies to non-GenICam devices, as can be seen
from comparing, for instance, \verb|qdet_peak| and \verb|qdet_eiger|
(plus their self-made SDKs) with their GenICam-like counterparts.

\begin{table}[htbp]\centering
\caption{%
	\prog{QDetectorIOC}'s \prog{ZeroMQ}-based data protocol;
	while its description involves interfaces of \prog{NumPy},
	this is purely meant to facilitate understanding, and the
	protocol itself can be implemented independent of Python.%
}\label{tbl:qdet-zmq}%\vspace*{1em}
\begin{minipage}[t]{0.95\textwidth}
\begin{enumerate}[itemsep = 0pt]
\item \prog{ZeroMQ} \verb|PUSH| sockets are used to output multidimensional
	data in QDetectorIOCs.  Receivers of the data streams bind to certain
	\prog{ZeroMQ} \verb|PULL| endpoints (\eg\ \verb|tcp://127.0.0.1:1234|),
	and let IOCs connect to these endpoints.
\item These \prog{ZeroMQ} endpoints are generally controlled
	with string PVs like \verb|${P}output0|, where \verb|${P}|
	is the PV prefix of the IOC.  PVs like \verb|${P}output1|
	are also provided by IOCs that can output multiple streams.
\item A multidimensional array is represented as a series of \prog{ZeroMQ}
	messages: a header message, zero to multiple data messages and
	an end message; to facilitate error handling, an end message with
	no corresponding header message or data messages is also allowed.
\item The end message is an empty non-multipart message.  The header
	message is a non-multipart message containing a JSON string.
	A data message should, in accordance with the header message,
	somehow encode one to multiple frames of data.
\item Encoded in the header message must be a dictionary that
	at least contains the fields \verb|shape| and \verb|dtype|,
	both of which should be values acceptable by \verb|numpy.frombuffer()|
	as the corresponding parameter: \eg\ \verb|shape| being
	\verb|[1080, 1920]| and \verb|dtype| being \verb|uint16|.
\item The header dictionary can have an optional \verb|variant| field
	which defaults to the empty string when unspecified, and perhaps
	other fields depending on the variant.  These fields are used to
	specify specialised formats for the following data messages.
\item When \verb|variant| is empty, every data message must be a non-multipart
	message, whose content can be passed to \verb|numpy.frombuffer()|
	with the current \verb|shape| and \verb|dtype|; of course, the
	former needs to be adapted for messages encoding multiple frames.
\end{enumerate}
\end{minipage}
\end{table}

In our summary, the performance limitations of \prog{areaDetector} and
its \prog{HDF5Plugin} (\cf\ \secref{intro}) mainly stem from two kinds of
performance overheads: \emph{per-frame overhead} and \emph{blocking overhead}.
The former stems from fragmented data passing: \eg\ each call to the
\verb|doCallbacksGenericPointer()| function in \prog{areaDetector} passes
one frame of data from device-specific readout code to plugins like
\prog{HDF5Plugin}, which incurs overhead that could be greatly reduced if
an interface was used that could pass a block composed of an indefinite
(but non-zero) number of frames.  The latter is a result of blocking
operations: \eg\ when doing data readout with \prog{Bluesky} \cite{allan2019},
\prog{HDF5Plugin} is by default set to a synchronous mode, which results in
every call to \verb|doCallbacksGenericPointer()| waiting for the completion
of the corresponding HDF5 writing operation.  With these two issues in mind,
we designed a data protocol for \prog{QDetectorIOC} (except for simple 0D
IOCs like \verb|qdet_nct| and \verb|qdet_o974|) based on asynchronous
transmission over \prog{ZeroMQ}, as is shown in \tblref{qdet-zmq}.  The
protocol is deliberately flexible: apart from its natural support for 0D/1D
transmission, by exploiting the \verb|variant| field in the header message,
compressed data (\eg\ those from Eiger) may also be transmitted, saving
bandwidth and leaving the work of decompression to downstream programs
like \prog{MDW} (\cf\ \secref{personnel}).  By additionally exploiting
variant-specific fields (\eg\ \verb|fields| when \verb|variant|
is set to \verb|table|), data like the ``tables'' from PandABox can
also be comfortably transmitted (\cf\ the \verb|qdet_panda| IOC).  We
would also like to note that \prog{QDetectorIOC} supports \emph{multiple
output streams} in one IOC, which currently seems hard to do systematically
in \prog{areaDetector}: \eg\ the \verb|uint32| and \verb|double| streams
from PandABox, as well as the image streams for two energy thresholds from
Eiger 2 when the latter's ``STREAM2'' interface \cite{burian2023} is enabled.

\prog{QDetectorIOC}'s data protocol is also intentionally simple to integrate.
On the production side, the protocol can be easily implemented on a per-device
basis in \prog{EPICS} IOCs; from our perspective, a realistic application
scenario for it is the readout of positions in each servo cycle from advanced
motion controllers (that support this function), which can be added to the
\prog{asynMotor} IOCs for these controllers.  On the consumption side, it can
also be easily supported by programs outside the \prog{MDW} framework; for
internal testing purposes, we have written real visualisation and data-saving
programs that use this protocol.  With our test program \verb|qpull.py|
(available from the supplementary materials), we have verified that the
\verb|qdet_panda| IOC can saturate the maximal data throughput of PandABox --
around 45\,MB/s using its \verb|XML FRAMED SCALED| format and 1\,Gb ethernet
-- which amounts to an about 0.98\,MHz framerate if each frame contained 6
\verb|double| fields.  By running \verb|qpull.py| on the 127.0.0.1 loopback
interface, we are able to verify that the maximal throughput of the Tucsen
camera in \tblref{perf-req} can also be saturated.  The Hamamatsu camera in
the same table has some issues that currently prevent data readout from its
CXP ports, and we will be able to test \verb|qdet_dcam|, our IOC for Hamamatsu
devices, against its maximal throughput after the issues get resolved.
Integration of devices with higher throughputs is already in progress,
and we are also investigating techniques like RDMA and multi-node readout
\cite{bideaud2020} for readout requirements on the 10\,GB/s level, aiming for
a simple, clear protocol based on thin, succinct encapsulation of the underlying
high-performance transport.  We are aware of the issue that Python is not a
performant language by itself, but we believe it is feasible to combine the
expressiveness of Python and the performance of C/C++ in data readout and
transmission: by passing relatively big blocks of data (reducing context-switch
overhead, just like the per-frame overhead above) from Python to C/C++, doing
high-performance processing on the C/C++ layer, and returning processed blocks
back to Python.  Actually, this is how high-performance Python libraries, \eg\ %
\prog{NumPy}/\prog{SciPy} and \prog{PyTorch}/\prog{TensorFlow}, are implemented.

\section{Conclusion}

The large number and strong diversity of detectors that need to be integrated
at HEPS and other light sources lead to the strong diversity in the software/%
hardware interfaces of these detectors.  By analysing the code of \prog{EPICS}
IOCs for detectors, we found what the device-specific code in them do can be
categorised into control communication and data communication.  A separation
of concerns is established at HEPS, that leaves the diversity in data
postprocessing to the \prog{MDW} framework, with a set of unified
data-transmission protocols being the boundary between it and us.  To avoid
an unlimited growth of workload due to the numerous control parameters
of most detectors, when doing detector integration we first focus on the
common, essential aspects of the control interfaces, and then help beamline
scientists ensure the correctness of more aspects according to feedback from
the latter.  With techniques like \prog{ihep-pkg} and \verb|~/iocBoot|,
we are able to minimise the workload in \prog{EPICS} deployment; similarly,
we make extensive use of \prog{ADAravis} to integrate most GigE/USB3 Vision
industrial cameras.  Following the same line of thought, we modified
\prog{ADGenICam} to allow for the unified integration, with \prog{EPICS}
IOCs based on it, of detectors conforming to the ``BCDEIS'' type system.

With the continuous advances in data throughputs and framerates of detectors,
mainstream software frameworks have themselves become a main bottleneck in the
application of high-performance detectors.  Noticing these issues, as well
as the code redundancy issue in \prog{areaDetector} and similar frameworks,
we created the \prog{QDetectorIOC} framework based on \prog{QueueIOC}.
By means of succinct self-made SDKs that expose consistent interfaces,
we are able to keep strong consistency between the code of QDetectorIOCs,
facilitating development and maintenance of the code.  Useful features in our
modified version of \prog{ADGenICam}, \eg\ the support for a JSON-based format
for device descriptions and the \verb|ADGenICamSlow| mechanism for updates
of feature subsets on certain conditions, are also cleanly reimplemented in
\prog{QDetectorIOC}.  \prog{QDetectorIOC} uses a versatile data protocol
based on \prog{ZeroMQ}, which naturally supports transmission of 0D/1D data;
it also supports specialised formats, \eg\ table-like data and compressed
data.  \prog{QDetectorIOC} can easily support multiple output streams; the
data protocol is also easy to integrate on both the production and consumption
sides.  Tests show that the protocol can saturate the 1.3\,GB/s data
throughput of a device, while integration of devices with higher throughputs
is already in progress; we are also investigating techniques like RDMA
and multi-node readout for readout requirements on the 10\,GB/s level.

\section*{Acknowledgements}

The authors would like to thank all beamlines at HEPS and BSRF for enabling
them to deal with the great variety of detectors discussed in this paper;
without the challenges and opportunities during this unique experience,
the paper would have never come into existence.  This work was supported
by the National Key Research and Development Program for Young Scientists
(Grant No.\ 2023YFA1609900) and the Young Scientists Fund of the National
Natural Science Foundation of China (Grants Nos.\ 12205328, 12305371).

\bibliography{art9}
\end{document}